\documentclass[11pt]{article}
\usepackage{amsmath,amssymb,amsthm,booktabs,array,graphicx}
\usepackage{algorithm,algpseudocode}
\usepackage[authoryear,round]{natbib}
\usepackage{hyperref}

\newtheorem{proposition}{Proposition}
\newtheorem{remark}{Remark}
\newtheorem{lemma}{Lemma}
\newcommand{\Cov}{\operatorname{Cov}}
\newcommand{\Var}{\operatorname{Var}}
\newcommand{\E}{\mathbb{E}}
\newcommand{\IF}{\widehat\Phi}
\title{Stable and efficient covariance estimation for sample L-moments using H\'ajek projections}
\author{\begin{tabular}{c}
Jihong Park$^{a}$\qquad Palakorn Seenoi$^{b,*}$\qquad
Jeong-Soo Park$^{c}$\\[7pt]
\small $^{a}$Department of Mathematics and Statistics, Chonnam National University\\
\small Gwangju 61186, South Korea\\[4pt]
\small $^{b}$Department of Statistics, Faculty of Science, Khon Kaen University\\
\small Khon Kaen 40002, Thailand\\[4pt]
\small $^{c}$Department of Statistics, Chonnam National University\\
\small Gwangju 61186, South Korea\\[5pt]
\small $^{*}$Corresponding author:
\href{mailto:palakorns@kku.ac.th}{palakorns@kku.ac.th}
\end{tabular}}
\date{}
\begin{document}
\maketitle
\begin{abstract}
L-moments are useful summary statistics and are widely used in statistical analysis.
The joint covariance matrix of sample L-moments not only describes their sampling
uncertainty but is also needed for covariance-weighted distances and related
statistical procedures. An exact unbiased covariance estimator is already
available. However, in our previous applications, this estimator sometimes
produced singular or indefinite covariance matrices. Such matrices cause problems when
matrix inversion is required. The present paper addresses this difficulty. We examine
standard positive-definite corrections and two covariance estimators based on
leave-one-out sample L-moments: the delete-one jackknife estimator and
an estimator based on the H\'ajek projection. Both are positive semidefinite by
construction.
Simulation results show that the projection estimator generally has smaller
estimation error and better numerical stability than the exact unbiased
estimator. The jackknife estimator overestimates the variances of higher-order sample
L-moments and has larger estimation error than the projection estimator. We also develop fast rank-prefix algorithms for the three
covariance estimators considered. These algorithms substantially
reduce computation time.
\end{abstract}
\noindent\textbf{Keywords:} covariance consistency; delete-one jackknife; eigenvalue correction;
Hoeffding decomposition; positive semidefiniteness; rank-prefix sums; U-statistics.
\section{Introduction}\label{sec:intro}
L-moments are widely used to describe the main features of probability
distributions. They are linear combinations of expected order statistics.
The first four L-moments describe location, scale, skewness, and kurtosis,
respectively. Their sample counterparts provide useful summary statistics.
In particular, the third and fourth L-moments, when divided by the second
L-moment, give the dimensionless measures called L-skewness and L-kurtosis.
These measures are useful for comparing samples and identifying suitable
distribution families. The so-called L-moment ratio diagram provides a
convenient graphical tool for these purposes
\citep{Hosking1990,VogelFennessey1993}.

L-moments also play an important role in statistical inference. They are used
in parameter and quantile estimation, regional frequency analysis, and
extreme-value modeling. Their advantages include simple computation and good
performance for small and moderate samples in many settings. Early work on
probability-weighted moments established useful estimators of generalized
extreme-value parameters and high quantiles \citep{HoskingWallisWood1985}.
More recent applications include threshold selection, confidence intervals
for extreme quantiles, regional frequency analysis, and L-moment-based
estimation for models of extreme values
\citep{SilvaLombaFragaAlves2020,Simkova2021,PrahadchaiBusababodhinPark2024,
BusababodhinShinPark2026}.

The sampling variability of L-moments is important for these applications.
When several L-moments are used together, their joint covariance matrix
describes both individual uncertainties and dependence among the estimates.
This matrix is needed for joint confidence regions and multivariate inference.
Applications involving L-moment-based confidence regions and statistical
inference can be found in \citet{Hosking2015}, \citet{WangHutson2013},
and \citet{BusababodhinShinPark2025}. Sampling uncertainty in L-moment-based
flood quantile estimation has also been studied by \citet{KjeldsenJones2004}.

\citet{Hosking2015} compared methods for constructing joint confidence
regions for L-skewness and L-kurtosis. He concluded that a bootstrap
method based on estimation of the covariance matrix of the sample L-moment
ratios is overall the best simple choice. He also found that the unbiased
estimator of \citet{ElamirSeheult2004} often gave invalid results for
$n<50$, because the estimated covariance matrix was not positive definite.
The bootstrap, however, recomputes the sample L-moments for each of $B$
resamples, with $B$ typically between 100 and 1000, so it costs about $B$
times as much as a single evaluation. The exact bootstrap covariance of
\citet{WangHutson2013} avoids resampling, but \citet{Hosking2015} found
it computationally taxing for $n>100$. The cost of the bootstrap becomes
substantial when
the covariance matrix must be estimated repeatedly, for example at many
sites in a regional analysis or within a Monte Carlo study. The bootstrap
estimate is also subject to Monte Carlo error. We therefore ask which
covariance estimator is fast to compute, deterministic, and positive
semidefinite while remaining accurate in finite samples. Constructing
confidence intervals or regions from such an estimator is a separate
problem and is not studied here.

Accurate estimation of this covariance matrix is therefore important. In
particular, some statistical procedures require its inverse, which in turn
requires a nonsingular and well-conditioned covariance estimate. An exact
unbiased estimator was developed by \citet{ElamirSeheult2004}; we call it
the ES estimator. Although unbiasedness
is an attractive theoretical property, it does not guarantee that the
estimated covariance matrix is positive semidefinite. While working on
related L-moment-based inference problems
\citep{ShinShinPark2026SERRA,ShinEtAl2026HSJ}, we encountered singular or
indefinite covariance estimates. These caused numerical difficulties when
the covariance matrix had to be inverted.

To address this problem, we consider two approaches. The first is an
eigenvalue-flooring correction of the exact unbiased covariance estimator.
This correction replaces small or negative eigenvalues with positive values.
It produces a positive definite matrix and controls its condition number.
Such eigenvalue corrections are well known in numerical matrix analysis
\citep{Higham1988,ChengHigham1998}. However, the correction changes the
original covariance estimate and may affect its statistical accuracy.

The second approach uses covariance estimators constructed from
leave-one-out sample L-moments, that is, from the sample L-moments computed
after one observation is removed. We consider two such estimators. The
first is based on the H\'ajek projection, a well-known method for
approximating a statistic by a sum of simpler terms
\citep{HajekSidakSen1999,OorschotSegersZhou2023}. We apply this projection
to sample L-moments and construct a covariance estimator from the resulting
approximation. We call it the HP estimator. The second is the
delete-one jackknife covariance estimator \citep{EfronStein1981,Efron1982},
which we call the JK estimator. The HP estimator coincides with the
structural-components variance estimator of \citet{Sen1960}, which is
widely used, for example, to compare areas under correlated ROC curves
\citep{DeLongEtAl1988}. Both estimators are positive semidefinite
by construction. Unlike eigenvalue flooring, they do not require a
correction of negative eigenvalues.

The two estimators are closely related. We show that the HP estimator is an
exact diagonal rescaling of the JK estimator. The two estimators give the
same variance estimate for the first L-moment. For higher-order L-moments,
the HP variance estimates are smaller than the JK variance estimates for
every sample. The two estimators also have the same first-order asymptotic
limit, but they can behave differently in finite samples. We therefore
include both estimators in our comparisons.

We examine the finite-sample properties of these approaches through Monte
Carlo simulations. We compare the ES estimator, its
eigenvalue-corrected versions, the HP estimator, and the JK estimator.
The comparisons focus on covariance estimation error, positive definiteness,
and numerical conditioning. We also investigate the finite-sample bias of
the HP and JK estimators. Our results show that the HP
estimator generally has smaller estimation error and better numerical
stability than the ES estimator in the settings examined. The JK estimator tends to
overestimate the variances of higher-order L-moments.

We also develop fast algorithms for computing the ES, HP, and JK
estimators. These algorithms use the so-called rank-prefix sums to avoid repeated calculations. This substantially reduces computation time.

A related projection estimator for the covariance of sample L-comoments
(multivariate L-moments),
together with exact covariance formulas and fast algorithms, is studied in
a companion paper \citep{SeenoiPark2026arXiv}. The present paper differs
in its focus. It concerns univariate sample L-moments, the definiteness
problem of the ES estimator, the exact relation between the HP and JK
estimators, and the comparison of these estimators with eigenvalue
flooring.

The remainder of the paper is organized as follows. Section~\ref{sec:estimators} introduces population and sample L-moments,
the ES, HP, and JK covariance estimators, and their asymptotic properties.
Section~\ref{sec:pd} studies positive-definite corrections and numerical
assessment. Section~\ref{sec:sim} presents simulation results.
Section~\ref{sec:fastHP} presents the fast rank-prefix algorithms and their
computing times. Sections~\ref{sec:discussion} and~\ref{sec:conclusion}
give discussion and conclusions. Technical details and pseudocode for the
fast algorithms are given in the Appendix.

\section{Covariance estimation for sample L-moments}\label{sec:estimators}
Let $X_1,\ldots,X_n$ be independent and identically distributed observations
from a population distribution $F$, where $\E(X^2)<\infty$.
Let $X_{1:n}\le\cdots\le X_{n:n}$ denote the order statistics of
$X_1,\ldots,X_n$. More generally, $X_{j:k}$ denotes the $j$th smallest
observation in a random sample of size $k$ from $F$.

\subsection{Population and sample L-moments}
The population L-moment of order $m$, denoted by $\lambda_m$, is defined as
\begin{equation}\label{eq:populationlmom}
\lambda_m=\frac{1}{m}\sum_{j=1}^{m}(-1)^{m-j}
\binom{m-1}{j-1}\E(X_{j:m}),\qquad m\ge 1.
\end{equation}
Here $\binom{m-1}{j-1}$ is a binomial coefficient. The first
L-moment is a measure of location, and the second is a measure of scale.
The third and fourth L-moments describe the shape of a distribution,
in particular its asymmetry and tail weight.
Specifically, $\lambda_1=\E(X)$ and
\begin{equation}\label{eq:ginilmom}
\lambda_2=\frac12\E(X_{2:2}-X_{1:2}).
\end{equation}
Thus, $2\lambda_2$ is Gini's mean difference, a familiar measure of
dispersion. For $\lambda_2>0$, the ratios
$\tau_3=\lambda_3/\lambda_2$ and $\tau_4=\lambda_4/\lambda_2$
are called L-skewness and L-kurtosis. They measure distributional
shape without depending on location or scale
\citep{HoskingWallis1997,VogelFennessey1993}.

The sample L-moment of order $m$, denoted by $l_m$, can be written as
a U-statistic:
\begin{equation}\label{eq:ustat}
l_m=\binom{n}{m}^{-1}\sum_{1\le i_1<\cdots<i_m\le n}
h_m(X_{i_1},\ldots,X_{i_m}),\qquad 1\le m\le n,
\end{equation}
where the symmetric kernel is
\begin{equation}\label{eq:kernel}
h_m(x_1,\ldots,x_m)=\frac1m\sum_{j=1}^m
(-1)^{m-j}\binom{m-1}{j-1}x_{j:m}.
\end{equation}
The U-statistic averages the kernel over all subsets of $m$ distinct
observations \citep{Hoeffding1948,Serfling1980,Hosking1990}.
Each subset in \eqref{eq:ustat} consists of $m$ independent observations
from $F$. Hence the ordered values in the kernel \eqref{eq:kernel} have
the same joint distribution as $X_{1:m},\ldots,X_{m:m}$, and comparing
\eqref{eq:kernel} with \eqref{eq:populationlmom} gives
$\E\{h_m(X_1,\ldots,X_m)\}=\lambda_m$. Because $l_m$ is an average of
such kernel values, linearity of expectation yields
\begin{equation}\label{eq:lmunbiased}
\E(l_m)=\lambda_m.
\end{equation}
Thus, each sample L-moment is unbiased for its population counterpart.
Moreover, if $\E(X^2)<\infty$, the sample L-moments are consistent for
the population L-moments, and any fixed finite vector of sample L-moments
is asymptotically normal \citep{Hosking1990}.

\subsection{Joint covariance and motivation}
Several L-moments are often used together, and their sampling errors
are generally correlated. Let $p$ denote the number of L-moments
under consideration, where $1\le p<n$. Define
$\boldsymbol l=(l_1,\ldots,l_p)^\top$ and
$\boldsymbol\lambda=(\lambda_1,\ldots,\lambda_p)^\top$.
Their finite-sample covariance matrix is
\begin{equation}\label{eq:targetcov}
\Sigma_n=\Cov(\boldsymbol l)
=[\Cov(l_m,l_q)]_{m,q=1}^p.
\end{equation}
The diagonal elements are the sampling variances of the individual
L-moments. The off-diagonal elements describe their dependence.
Although $\Sigma_n$ is the covariance matrix of the random vector
$\boldsymbol l$, it is itself a fixed, nonrandom matrix that depends only on
$F$ and $n$. In this study we mainly consider $p=4$.

Asymptotic covariance formulas are useful for large samples.
However, their accuracy can be limited for small and moderate samples,
as often encountered in hydrological and extreme-value studies.
Reliable covariance estimation is therefore important for various
purposes of finite-sample statistical inference, such as joint confidence
regions, hypothesis tests, and model comparison
\citep{ElamirSeheult2004,Hosking2015}.
L-moment-based statistical distances also arise in estimation and
model comparison \citep{ShinEtAl2026HSJ}.
\citet{KjeldsenJones2004} studied the sampling variance of flood
quantiles estimated by the method of L-moments. Such a variance depends on
the covariance matrix of the sample L-moments, so that an accurate and
efficient estimator of this matrix is needed in practice.

An exact distribution-free unbiased covariance estimator was developed
by \citet{ElamirSeheult2004}. As in Section~\ref{sec:intro}, we call it the
Elamir--Seheult (ES) estimator. Its estimated covariance matrix can,
however, be singular or indefinite. We therefore consider an estimator
based on the H\'ajek projection, called the HP estimator, and the
delete-one jackknife estimator, called the JK estimator, as well as
positive-definite corrections of ES. In comparing these estimators, we
focus on their finite-sample accuracy, matrix definiteness, and
computational efficiency.

\subsection{Exact unbiased covariance estimator}
The ES estimator is based on the unbiased sample
probability-weighted moments (PWMs),
\begin{equation}\label{eq:pwm}
b_k=\frac1n\sum_{i=1}^n\frac{(i-1)_{(k)}}{(n-1)_{(k)}}X_{i:n},
\qquad k=0,1,\ldots,p-1,
\end{equation}
where $(a)_{(r)}=a(a-1)\cdots(a-r+1)$, $(a)_{(0)}=1$, and
$(a)_{(r)}=0$ when $a$ is a nonnegative integer smaller than $r$.
The sample L-moments are linear combinations of the PWMs,
\begin{equation}\label{eq:lCb}
\boldsymbol l=C\boldsymbol b,\qquad
C_{m,k+1}=(-1)^{m-1-k}\binom{m-1}{k}\binom{m-1+k}{k},
\end{equation}
where $\boldsymbol b=(b_0,\ldots,b_{p-1})^\top$ and $C$ is the $p\times p$
matrix with the entries above for $0\le k\le m-1$ and zero otherwise
\citep{Hosking1990}.
\citet{ElamirSeheult2004} showed that
\begin{equation}\label{eq:esb}
\widehat{\Cov}(b_k,b_l)=b_kb_l-\frac{1}{(n)_{(k+l+2)}}
\sum_{1\le i<j\le n}\bigl\{(i-1)_{(k)}(j-k-2)_{(l)}
+(i-1)_{(l)}(j-l-2)_{(k)}\bigr\}X_{i:n}X_{j:n}
\end{equation}
is an unbiased estimator of $\Cov(b_k,b_l)$ for $0\le k,l\le p-1$ and
$n\ge2p$. Let $\widehat S_b$ be the $p\times p$ matrix with entries
\eqref{eq:esb}. The ES estimator of $\Sigma_n$ is
\begin{equation}\label{eq:es}
\widehat\Sigma_{\mathrm{ES}}=C\widehat S_bC^\top ,
\end{equation}
and it satisfies
\begin{equation}\label{eq:unbiased}
\E(\widehat\Sigma_{\mathrm{ES}})=\Sigma_n.
\end{equation}
We compute it with the \texttt{lmoms.cov} function in the
\texttt{lmomco} package \citep{Asquith2026}. The same estimator is also
available as the \texttt{Lmomcov} function in the \texttt{Lmoments} package
\citep{KarvanenKarppinen2025} and as the \texttt{varLmoments} function in
the \texttt{nsRFA} package \citep{Viglione2024}. Appendix~\ref{app:fast}
gives a faster algorithm for the same estimator.

\subsection{Contributions, the H\'ajek projection, and the influence function}\label{sec:hp}
We now derive the HP estimator in two steps. This subsection describes how a
single observation contributes to the sample L-moment $l_m$ and expresses
these contributions through the influence function.
Section~\ref{sec:hpest} replaces the unknown influence values by empirical
ones and forms their sample covariance matrix. Detailed derivations are
given in Appendix~\ref{app:supp}.

Recall that $l_m$ is a U-statistic with symmetric kernel $h_m$ and
expectation $\lambda_m$. Define
\begin{equation}\label{eq:gm}
g_m(x)=\E\{h_m(x,X_2,\ldots,X_m)\}-\lambda_m.
\end{equation}
The expectation is taken over the remaining $m-1$ independent
observations. Thus, $g_m(x)$ describes how the expected kernel changes
when one argument is fixed at $x$, and $\E\{g_m(X)\}=0$.
We define the \emph{contribution} of observation $i$ to $l_m$ as
\begin{equation}\label{eq:contribdef}
c_{m,i}=\E(l_m-\lambda_m\mid X_i)=\frac mn\,g_m(X_i);
\end{equation}
the second equality is shown in Appendix~\ref{app:suppproj}. Each
contribution has mean zero, as does $l_m-\lambda_m$.

The \emph{H\'ajek projection} of $l_m-\lambda_m$ is the sum of the
contributions \citep{Hajek1968,HajekSidakSen1999},
\begin{equation}\label{eq:hajekprojection}
H_{n,m}=\sum_{i=1}^nc_{m,i}=\frac{m}{n}\sum_{i=1}^n g_m(X_i).
\end{equation}
It is the $L^2$ projection of $l_m-\lambda_m$, that is, its closest
approximation in mean square, onto the space of sums
$\sum_{i=1}^na_i(X_i)$ of square-integrable functions of single
observations (Appendix~\ref{app:suppproj}).

More generally, the
\emph{Hoeffding decomposition} \citep{Hoeffding1948,Lee1990} writes
$l_m-\lambda_m$ as a sum of mutually uncorrelated terms: the $n$
single-observation terms $c_{m,i}$, terms that depend jointly on pairs of
observations, and so on, up to terms that depend on $m$ observations. A
pair term depends on $X_i$ and $X_j$ together, and it cannot be split into
a part due to $X_i$ alone and a part due to $X_j$ alone. These interaction
terms are not part of any $c_{m,i}$. Their sum is the remainder $R_{n,m}$
in \eqref{eq:hoeffdecomp} of Section~\ref{sec:asymptotics}.

Next, define
\begin{equation}\label{eq:phi}
\phi_m(x)=m g_m(x).
\end{equation}
Substituting \eqref{eq:phi} into \eqref{eq:hajekprojection} gives
$H_{n,m}=n^{-1}\sum_{i=1}^n\phi_m(X_i)$. Thus $H_{n,m}$ is an average of
$n$ independent mean-zero terms, and
$\Var(H_{n,m})=n^{-1}\E\{\phi_m(X)^2\}$. This form is used in
Section~\ref{sec:asymptotics}.

The function $\phi_m$ is the
\emph{influence function} of $\lambda_m$. In its original sense, the
influence function is the derivative of a functional under
infinitesimal contamination of $F$ at $x$ \citep{Serfling1980,vanDerVaart1998}.
Write $\lambda_m(F)=\E_F\{h_m(X_1,\ldots,X_m)\}$, where $\E_F$ denotes
expectation when $X_1,\ldots,X_m$ are i.i.d.\ with distribution $F$, and let
$\Delta_x$ be the distribution with unit point mass at $x$. Then
\begin{equation}\label{eq:ifderiv}
\operatorname{IF}_m(x;F)
=\left.\frac{d}{d\varepsilon}\,
\lambda_m\big((1-\varepsilon)F+\varepsilon\Delta_x\big)
\right|_{\varepsilon=0}
=m\,g_m(x)=\phi_m(x);
\end{equation}
see Appendix~\ref{app:suppif} for the second equality.

The contribution
and the influence function are therefore related by
\begin{equation}\label{eq:contribif}
c_{m,i}=\frac1n\,\phi_m(X_i).
\end{equation}
That is, the contribution of an observation is its influence divided by the
sample size. We call $\phi_m(X_i)=n\,c_{m,i}$ the \emph{influence
contribution} of observation $i$; it is the contribution expressed on the
scale of a single observation.

\subsection{Empirical influence values and the HP estimator}\label{sec:hpest}
The functions $g_m$ and $\phi_m$ depend on the unknown distribution $F$.
We therefore use sample-based (empirical) influence values. Let
$l_m^{(-i)}$ denote the sample L-moment computed after omitting $X_i$.
Define
\begin{equation}\label{eq:leaveout}
\widehat\phi_{m,i}=(n-m)\{l_m-l_m^{(-i)}\},\qquad n>m.
\end{equation}
This is the change in the sample L-moment caused by removing $X_i$,
multiplied by $n-m$. The factor $n-m$ makes $\widehat\phi_{m,i}$ an
estimate of $\phi_m(X_i)$, as the following identity shows. Let
$A_{m,i}$ denote the average
of $h_m$ over all subsets of size $m$ that contain observation $i$.
Counting these subsets gives
\begin{equation}\label{eq:subsetcount}
l_m=\frac{m}{n}A_{m,i}+\frac{n-m}{n}l_m^{(-i)}.
\end{equation}
Consequently,
\begin{equation}\label{eq:contribution}
\widehat\phi_{m,i}=m(A_{m,i}-l_m).
\end{equation}
Since $\phi_m(x)=m\{\E h_m(x,X_2,\ldots,X_m)-\lambda_m\}$,
\eqref{eq:contribution} is its empirical counterpart: the expectation is
replaced by the average $A_{m,i}$, and $\lambda_m$ by $l_m$.
The subsets that contain $i$ also carry the interaction terms that involve
$X_i$, so $\widehat\phi_{m,i}$ is an estimate of $\phi_m(X_i)$, not an exact
separation of the part of $l_m$ that is due to $X_i$ alone. Indeed,
\begin{equation}\label{eq:condexp}
\E(\widehat\phi_{m,i}\mid X_i)=\frac{n-m}{n}\,\phi_m(X_i)
\end{equation}
(Appendix~\ref{app:suppemp}), so $\widehat\phi_{m,i}$ is conditionally
shrunk toward zero by the factor $(n-m)/n$, which tends to one as
$n\to\infty$. The empirical influence values are also automatically
centered (Appendix~\ref{app:suppemp}):
\begin{equation}\label{eq:center}
\sum_{i=1}^n\widehat\phi_{m,i}=0.
\end{equation}

By \eqref{eq:contribif}, the H\'ajek projections have covariances
\[
\Cov(H_{n,m},H_{n,q})=\frac1n\E\{\phi_m(X)\phi_q(X)\}.
\]
Replacing this expectation by the sample covariance of the empirical
influence values gives the HP estimator. Let $\IF$ be the $n\times p$
matrix whose entry in row $i$ and column $m$ is $\widehat\phi_{m,i}$. We
define the HP covariance estimator by
\begin{equation}\label{eq:hp}
\widehat\Sigma_{\mathrm{HP}}=\frac{1}{n(n-1)}\IF^\top\IF.
\end{equation}
It is the sample covariance matrix of empirical influence vectors,
divided by $n$; no centering is needed because of \eqref{eq:center}. By
\eqref{eq:contribution}, its $(m,q)$ entry is
$mq\{n(n-1)\}^{-1}\sum_{i=1}^n(A_{m,i}-l_m)(A_{q,i}-l_q)$, which is the
multivariate form of the structural-components variance estimator of
\citet{Sen1960}. For any $a\in\mathbb R^p$,
\begin{equation}\label{eq:hppsd}
a^\top\widehat\Sigma_{\mathrm{HP}}a
=\frac{\|\IF a\|^2}{n(n-1)}\ge0.
\end{equation}
Hence the HP estimate is positive semidefinite. It need not be
positive definite if the empirical influence vectors are linearly
dependent. The related projection estimator for sample L-comoments is
studied by \citet{SeenoiPark2026arXiv}.

\subsection{Delete-one jackknife covariance estimator}\label{sec:jk}
The leave-one-out sample L-moments $l_m^{(-i)}$ in \eqref{eq:leaveout}
also define the delete-one jackknife covariance estimator
\citep{EfronStein1981,Efron1982}. Let
$\boldsymbol l^{(-i)}=(l_1^{(-i)},\ldots,l_p^{(-i)})^\top$ and
$\bar{\boldsymbol l}^{(\cdot)}=n^{-1}\sum_{i=1}^n\boldsymbol l^{(-i)}$.
The delete-one jackknife (JK) covariance estimator is
\begin{equation}\label{eq:jk}
\widehat\Sigma_{\mathrm{JK}}=\frac{n-1}{n}\sum_{i=1}^n
\bigl(\boldsymbol l^{(-i)}-\bar{\boldsymbol l}^{(\cdot)}\bigr)
\bigl(\boldsymbol l^{(-i)}-\bar{\boldsymbol l}^{(\cdot)}\bigr)^\top .
\end{equation}
It equals the sample covariance matrix of the pseudo-values
$n\boldsymbol l-(n-1)\boldsymbol l^{(-i)}$, $i=1,\ldots,n$, divided by $n$
\citep{Efron1982}. For any $a\in\mathbb R^p$,
\begin{equation}\label{eq:jkpsd}
a^\top\widehat\Sigma_{\mathrm{JK}}a
=\frac{n-1}{n}\sum_{i=1}^n
\bigl\{a^\top\bigl(\boldsymbol l^{(-i)}-\bar{\boldsymbol l}^{(\cdot)}\bigr)\bigr\}^2\ge0 .
\end{equation}
Hence the JK estimate is also positive semidefinite by construction.

We now derive the relation between the JK and HP estimators. By
\eqref{eq:leaveout} and \eqref{eq:center},
\[
\sum_{i=1}^n(n-m)\{l_m-l_m^{(-i)}\}=0,
\]
so that
\begin{equation}\label{eq:loomean}
\frac1n\sum_{i=1}^n l_m^{(-i)}=l_m,\qquad m=1,\ldots,p .
\end{equation}
Thus $\bar{\boldsymbol l}^{(\cdot)}=\boldsymbol l$, and
$\widehat\phi_{m,i}=-(n-m)\{l_m^{(-i)}-\bar l_m^{(\cdot)}\}$.
Substituting this expression into \eqref{eq:hp} gives
\[
(\widehat\Sigma_{\mathrm{HP}})_{mq}
=\frac{(n-m)(n-q)}{n(n-1)}\sum_{i=1}^n
\{l_m^{(-i)}-\bar l_m^{(\cdot)}\}\{l_q^{(-i)}-\bar l_q^{(\cdot)}\}
=\frac{(n-m)(n-q)}{(n-1)^2}(\widehat\Sigma_{\mathrm{JK}})_{mq}.
\]
In matrix form,
\begin{equation}\label{eq:hpjk}
\widehat\Sigma_{\mathrm{HP}}=D_n\widehat\Sigma_{\mathrm{JK}}D_n,\qquad
D_n=\operatorname{diag}\Bigl(\frac{n-1}{n-1},\frac{n-2}{n-1},\ldots,
\frac{n-p}{n-1}\Bigr).
\end{equation}
The relation holds exactly for every sample. It has a simple
interpretation. The jackknife uses the generic pseudo-value scaling
$(n-1)\{l_m-l_m^{(-i)}\}$, which is appropriate for a general smooth
statistic. The HP estimator uses the factor $(n-m)$, which follows from
the subset count \eqref{eq:subsetcount} for a U-statistic of degree $m$.

The relation \eqref{eq:hpjk} gives the following properties of the JK
estimator.
\begin{enumerate}
\item The two estimators give the same variance estimate for the first
L-moment. This common
estimate is $s^2/n$ for $l_1=\bar X=n^{-1}\sum_{i=1}^nX_i$, where $s^2$ is
the sample variance.
The off-diagonal elements $(1,q)$ with $q\ge2$ differ by the factor
$(n-q)/(n-1)$.
\item Write $\widehat V_{M,m}=(\widehat\Sigma_M)_{mm}$ for the $m$th diagonal
element of an estimate $\widehat\Sigma_M$, where $M$ is ES, HP, or JK.
Because $0<(n-m)/(n-1)\le1$, the diagonal elements satisfy
\begin{equation}\label{eq:jkdiag}
\widehat V_{\mathrm{HP},m}=\Bigl(\frac{n-m}{n-1}\Bigr)^2
\widehat V_{\mathrm{JK},m}\le\widehat V_{\mathrm{JK},m}
\end{equation}
for every sample, with strict inequality for $m\ge2$ whenever
$\widehat V_{\mathrm{JK},m}>0$. For $n=20$ the multipliers are
$0.898$, $0.801$, and $0.709$ for $m=2,3,4$. For $n=50$ the multipliers
are $0.960$, $0.920$, and $0.881$.
\item Since $D_n$ is nonsingular for $n>p$, the two estimates have the
same rank. Hence $\widehat\Sigma_{\mathrm{JK}}$ is positive definite if and
only if $\widehat\Sigma_{\mathrm{HP}}$ is positive definite. Their condition
numbers can differ, but, with $\kappa_2(\cdot)$ denoting the spectral
condition number defined in \eqref{eq:cond},
$\kappa_2(\widehat\Sigma_{\mathrm{HP}})\le\{(n-1)/(n-p)\}^2
\kappa_2(\widehat\Sigma_{\mathrm{JK}})$ and the reverse inequality also
holds with the same factor.
\item For fixed $p$, $D_n\to I_p$ as $n\to\infty$. Hence, by
\eqref{eq:hpjk}, $n\widehat\Sigma_{\mathrm{JK}}$ and $n\widehat\Sigma_{\mathrm{HP}}$
have the same probability limit, and
$n(\widehat\Sigma_{\mathrm{JK}}-\widehat\Sigma_{\mathrm{HP}})\xrightarrow{p}0$.
Appendix~\ref{app:asymptotics} shows that this limit is the asymptotic
covariance matrix $\Omega$ of $\sqrt n(\boldsymbol l-\boldsymbol\lambda)$,
defined in \eqref{eq:omega} of Section~\ref{sec:asymptotics}, which is
also the limit for the ES estimator (Proposition~\ref{prop:equiv}).
\end{enumerate}
Neither estimator is unbiased for $\Sigma_n$. Their diagonal elements,
however, satisfy the following bounds. Let $S$ be a symmetric statistic of
$n-1$ i.i.d.\ observations, and let $S_{(i)}$ be its value on the $n-1$
observations that remain after $X_i$ is removed from $X_1,\ldots,X_n$.
\citet{EfronStein1981} showed that
$\E\{\sum_{i=1}^n(S_{(i)}-\bar S_{(\cdot)})^2\}\ge\Var(S)$, where
$\bar S_{(\cdot)}=n^{-1}\sum_{i=1}^n S_{(i)}$. Combined with a monotonicity
property of U-statistic variances, this gives the following result.

\begin{proposition}[Bias direction]\label{prop:bias}
Suppose the observations are i.i.d.\ with $\E(X^2)<\infty$ and $n>p$. Then,
for $m=1,\ldots,p$,
\[
\E(\widehat V_{\mathrm{JK},m})\ge\Var(l_m),\qquad
\E(\widehat V_{\mathrm{HP},m})\ge\Bigl(\frac{n-m}{n-1}\Bigr)^2\Var(l_m).
\]
\end{proposition}
\begin{proof}
See Appendix~\ref{app:hpjackknife}.
\end{proof}
Thus the JK variance estimates are conservative. The bound for the HP
estimator does not determine the sign of its bias, which is examined
empirically in Section~\ref{sec:bias}.
Both estimators are computed from the same $n$ leave-one-out vectors
$\boldsymbol l^{(-i)}$. Therefore the fast algorithm in
Appendix~\ref{app:fast} applies to both, and their computational costs are
the same.

\subsection{First-order asymptotics and equivalence}\label{sec:asymptotics}
We now describe the asymptotic sampling covariance of sample L-moments.
We then compare the limits of the ES, HP, and JK covariance estimators.
Suppose $p$ is fixed as $n\to\infty$, the observations are i.i.d.,
and $\E(X^2)<\infty$. These conditions ensure that the L-moment
kernels of each fixed degree have finite second moments
\citep{Hoeffding1948,Serfling1980,Lee1990}.

Recall that the H\'ajek projection is
$H_{n,m}=n^{-1}\sum_i\phi_m(X_i)$.
The Hoeffding decomposition expresses the centered U-statistic as
\begin{equation}\label{eq:hoeffdecomp}
l_m-\lambda_m=H_{n,m}+R_{n,m},
\end{equation}
where $R_{n,m}$ is the sum of higher-order terms. The distinct
components in this decomposition are mutually orthogonal
(Section~\ref{sec:hp}). Hence
\begin{equation}\label{eq:orthvar}
\Var(l_m)=\Var(H_{n,m})+\Var(R_{n,m}),\qquad
\Var(H_{n,m})=\frac1n\E\{\phi_m(X)^2\}.
\end{equation}
In particular, $\Var(H_{n,m})\le\Var(l_m)$. For fixed $m$, the
remainder satisfies $\E(R_{n,m}^2)=O(n^{-2})$.

Set $\boldsymbol\phi(x)=(\phi_1(x),\ldots,\phi_p(x))^\top$.
It follows that
\begin{equation}\label{eq:asylin}
\sqrt n(\boldsymbol l-\boldsymbol\lambda)
=\frac1{\sqrt n}\sum_{i=1}^n\boldsymbol\phi(X_i)+o_p(1).
\end{equation}
The asymptotic covariance matrix is
\begin{equation}\label{eq:omega}
\Omega=\E\{\boldsymbol\phi(X)\boldsymbol\phi(X)^\top\},\qquad
\Omega_{mq}=mq\,\Cov\{g_m(X),g_q(X)\}.
\end{equation}
Thus $n\Sigma_n\to\Omega$ and
$\sqrt n(\boldsymbol l-\boldsymbol\lambda)\xrightarrow{d}
N_p(\boldsymbol0,\Omega)$ \citep{Hoeffding1948,Hosking1990}; see also
Appendix~\ref{app:asymptotics}. Here $\Sigma_n$ in \eqref{eq:targetcov}
is the exact covariance matrix of $\boldsymbol l$ for sample size $n$,
whereas $\Omega$ does not depend on $n$ and is the asymptotic covariance
matrix of $\sqrt n(\boldsymbol l-\boldsymbol\lambda)$. Since $\Sigma_n$ is
not random, $n\Sigma_n\to\Omega$ is an ordinary limit. The matrix
$\Omega$ may be singular.
Proposition~\ref{prop:equiv} shows that the ES, HP, and JK covariance
estimators, multiplied by $n$, all converge to this limit.

\begin{proposition}[First-order equivalence]\label{prop:equiv}
Suppose $p$ is fixed, the observations are i.i.d. with
$\E(X^2)<\infty$, and $n\ge2p$. Then
\begin{equation}\label{eq:equiv}
n\widehat\Sigma_{\mathrm{ES}}\xrightarrow{p}\Omega,\qquad
n\widehat\Sigma_{\mathrm{HP}}\xrightarrow{p}\Omega,\qquad
n\widehat\Sigma_{\mathrm{JK}}\xrightarrow{p}\Omega,
\end{equation}
and hence
\[
n(\widehat\Sigma_{\mathrm{ES}}-\widehat\Sigma_{\mathrm{HP}})\xrightarrow{p}0,\qquad
n(\widehat\Sigma_{\mathrm{ES}}-\widehat\Sigma_{\mathrm{JK}})\xrightarrow{p}0,\qquad
n(\widehat\Sigma_{\mathrm{HP}}-\widehat\Sigma_{\mathrm{JK}})\xrightarrow{p}0.
\]
The convergence is entrywise and hence in every matrix norm.
\end{proposition}
\begin{proof}
The proof, given in Appendix~\ref{app:asymptotics}, uses the
finite-overlap representation of U-statistic covariances for the ES
estimator, and the consistency of the jackknife for U-statistics
\citep{Arvesen1969} for the JK and HP estimators.
\end{proof}
The condition $n\ge2p$ allows disjoint subsets in that representation.
For $p=4$ it requires $n\ge8$, a mild condition for the sample sizes
considered here. The proposition does not imply the same finite-sample
bias, estimation error, or numerical conditioning.

\section{Positive-definite correction and numerical assessment}\label{sec:pd}
\subsection{Eigenvalue flooring}
For a symmetric estimate $S$, let
$S=Q\operatorname{diag}(d_1,\ldots,d_p)Q^\top$ be its spectral
(eigenvalue) decomposition. Here $d_1,\ldots,d_p$ are the eigenvalues of
$S$, and the columns of the orthogonal matrix $Q$ are the corresponding
eigenvectors. Let $d_{\max}=\max_jd_j$ and $d_{\min}=\min_jd_j$.
Given $\delta>0$, define
\begin{equation}\label{eq:floor}
\mathcal P_\delta(S)=Q\operatorname{diag}\{\max(d_j,\delta):1\le j\le p\}Q^\top.
\end{equation}
Thus $\mathcal P_\delta(S)$ keeps the eigenvectors of $S$ and raises every
eigenvalue smaller than $\delta$ to $\delta$. For a prespecified positive
threshold $\delta$, the transformation in \eqref{eq:floor} uniquely
minimizes $\|T-S\|_F$ over symmetric $T\succeq\delta I_p$, a spectral
extension of the nearest-PSD principle \citep{Higham1988}. Here
$\|A\|_F=\{\sum_{j,k}a_{jk}^2\}^{1/2}$ is the Frobenius norm, and
$T\succeq\delta I_p$ means that $T-\delta I_p$ is positive semidefinite,
that is, every eigenvalue of $T$ is at least $\delta$. This is a statement about distance from the original matrix.
It does \emph{not} say that the correction minimizes MSE for
estimating $\Sigma_n$.

The threshold $\delta$ is chosen through the condition number of the
corrected matrix. For a positive definite matrix $S$, the spectral
condition number is
\begin{equation}\label{eq:cond}
\kappa_2(S)=d_{\max}/d_{\min},
\end{equation}
the ratio of the largest to the smallest eigenvalue. It is the
condition number with respect to the spectral (matrix 2-) norm, and for a
symmetric positive definite matrix it coincides with the usual condition
number $\|S\|_2\|S^{-1}\|_2$. A large value indicates that $S$ is close to
singular and that its inverse is sensitive to small perturbations.
Throughout the paper, ``condition number'' means this spectral
condition number $\kappa_2$.

We choose $\delta$ so that the corrected matrix satisfies
$\kappa_2\{\mathcal P_\delta(S)\}\le K$ for a prespecified upper bound
$K>1$, which we call the condition-number cap. We investigate
$K\in\{10^2,10^3,10^5\}$ and use $K=10^3$ as the default. This value
exceeded the condition number of the reference covariance matrix in five
of the six simulation settings (Table~\ref{tab:cond}).
Bounding the condition number of a covariance estimate in this way is a
common device for obtaining a numerically stable estimate; for example,
\citet{WonEtAl2013} impose such a bound in maximum likelihood estimation
of a covariance matrix. For a positive definite $S$, the simplest choice
is $\delta=d_{\max}/K$, which raises every eigenvalue below $d_{\max}/K$
to that value \citep{Higham1988,WonEtAl2013}. For an indefinite or nearly
singular $S$, we use the following modification of this rule:
\begin{equation}\label{eq:delta}
 s=\max_j|d_j|,\qquad
 u=\max(d_{\max},s/K),\qquad
 \delta=u/K.
\end{equation}
The implementation uses $s=1$ only in the exceptional zero-matrix case;
otherwise \eqref{eq:delta} follows directly. The corrected eigenvalues are
$d_j^*=\max(d_j,\delta)$. If $d_{\max}>0$ and $s=d_{\max}$, this reduces
to the familiar $\delta=d_{\max}/K$. The more general safeguard accommodates
an estimate having predominantly negative eigenvalues.

We use this relative threshold rather than a fixed absolute one, such as
$\delta=10^{-8}$. The entries of a covariance estimate scale with the
square of the measurement unit of the data, so a fixed threshold may be
negligible for one data set and dominant for a rescaled version of the
same data. The relative threshold satisfies
$\mathcal P_{\delta(cS)}(cS)=c\,\mathcal P_{\delta(S)}(S)$ for every
$c>0$, and it directly controls the condition number of the corrected
matrix. With the numerical
safeguard and $K>1$, the resulting matrix is PD and has spectral condition
number at most $K$, modulo floating-point arithmetic. Only the ES
estimator is corrected in the present simulation; the HP and JK estimators
are kept unmodified to show their intrinsic behavior.

\subsection{Definiteness versus sensitivity to inversion}
Let $d_{\min}$ be the smallest eigenvalue of the symmetrized estimate
$(S+S^\top)/2$ (see Appendix~\ref{app:alg}) and set, for a prespecified $\varepsilon_{\rm rel}>0$,
\begin{equation}\label{eq:tol}
\tau=\varepsilon_{\rm rel}\max\{\max_j|d_j|,\epsilon_{\rm mach}\},
\qquad \varepsilon_{\rm rel}=10^{-10},
\end{equation}
where $\epsilon_{\rm mach}\approx2.2\times10^{-16}$ is the machine epsilon of
double-precision arithmetic. For example, the largest eigenvalue of the
reference covariance matrix is about $0.050$ in the normal case with
$n=20$ and about $0.234$ in the GPD case with $n=20$, so that
$\tau\approx5\times10^{-12}$ and $\tau\approx2.3\times10^{-11}$,
respectively.
We classify an estimate as follows:
\begin{enumerate}
\item \emph{numerically non-PSD} when $d_{\min}<-\tau$;
\item \emph{near-singular} when $|d_{\min}|\le\tau$;
\item \emph{numerically PD} when $d_{\min}>\tau$.
\end{enumerate} The non-PD rate combines the first two categories.
The small machine-epsilon term prevents a zero threshold for the zero matrix
and matches the computation. Because the tolerance is scale-relative, this
classification differs from exact mathematical definiteness near zero.

For a numerically PD matrix, we report the spectral condition number
$\kappa_2(S)$ in \eqref{eq:cond}. We use $\kappa_2(S)$ as an indicator of numerical sensitivity during
matrix inversion. A small condition number is helpful. It does not
prove that a particular computer algorithm is numerically stable. Condition numbers are not direct measures
of statistical accuracy, because the population covariance matrix
$\Sigma_n$ can itself be ill-conditioned. A smaller value of
$\kappa_2(S)$ always makes the inversion of $S$ less sensitive to
numerical errors. For statistical accuracy, however, the relevant
comparison is with $\kappa_2(\Sigma_n)$. An estimate whose condition
number is much larger than $\kappa_2(\Sigma_n)$ is unnecessarily close to
singular. Conversely, flooring with a small cap $K$ can make the condition number
of an estimate much smaller than $\kappa_2(\Sigma_n)$. In that case the
small eigenvalues of the estimate are raised far above the small
eigenvalues of $\Sigma_n$, so the estimate no longer represents the
directions of small variability correctly, even though it is easy to
invert. Thus, when $\Sigma_n$ is
well-conditioned, a small $\kappa_2(S)$ is desirable, but a smaller
condition number does not by itself indicate a more accurate
estimate. Appendix~\ref{app:alg} gives an explicit classification
algorithm and correction pseudocode.

\section{Monte Carlo experiments}\label{sec:sim}
\subsection{Design and reference covariance}
We evaluate the four sample L-moments ($p=4$) under three sampling
distributions: standard normal $N(0,1)$; standard Gumbel with
$F(x)=\exp\{-\exp(-x)\}$; and the generalized Pareto distribution (GPD)
with scale one and shape
$\xi=0.2$, with quantile $F^{-1}(u)=((1-u)^{-\xi}-1)/\xi$.
Sample sizes are $n\in\{20,50\}$, with $B=1000$ independent evaluation
replications per setting. The reference finite-sample covariance
$\Sigma_n$ is approximated independently from $N_{\rm ref}=100000$
replications of the sample L-moment vector. Reference replicates are
separate from evaluation replicates; each estimate is compared to the
same setting-specific reference $\widehat\Sigma_{\rm ref}$.
Reference matrices were saved for reuse with checked simulation metadata.

We compare $\mathrm{ES}=\widehat\Sigma_{\mathrm{ES}}$,
$\mathrm{HP}=\widehat\Sigma_{\mathrm{HP}}$,
$\mathrm{JK}=\widehat\Sigma_{\mathrm{JK}}$, and
$\mathrm{ES}_K=\mathcal P_{\delta(K)}(\widehat\Sigma_{\mathrm{ES}})$ for
$K=100$, $1000$, and $100000$. Here $\mathrm{ES}_K$ is the ES estimate after
eigenvalue flooring with condition-number cap $K$ (Section~\ref{sec:pd}).
For an estimate $S_b$ in replication $b$, we call
$\|S_b-\widehat\Sigma_{\rm ref}\|_F^2$ its squared Frobenius error, where
$\|\cdot\|_F$ is the Frobenius norm defined in Section~\ref{sec:pd}. The
relative root mean squared Frobenius error (RRMSE) is
\begin{equation}\label{eq:rrmse}
\operatorname{RRMSE}_F(S)=
\frac{\{B^{-1}\sum_{b=1}^{B}\|S_b-\widehat\Sigma_{\rm ref}\|_F^2\}^{1/2}}
{\|\widehat\Sigma_{\rm ref}\|_F}.
\end{equation}
Because all methods are applied to the same samples, we also compare
method $M$ with HP sample by sample, through the paired difference
\begin{equation}\label{eq:paired}
D_b(M)=\|S_{M,b}-\widehat\Sigma_{\rm ref}\|_F^2
-\|S_{HP,b}-\widehat\Sigma_{\rm ref}\|_F^2,\quad
\widehat{\operatorname{MCSE}}(\overline D)=s_D/\sqrt B.
\end{equation}
Here $s_D$ is the sample standard deviation of $D_1(M),\ldots,D_B(M)$,
and MCSE denotes the Monte Carlo standard error. A positive mean
$\overline D$ means that HP has the smaller squared Frobenius error on
average. We report approximate 95\% intervals
$\overline D\pm1.96\,\widehat{\operatorname{MCSE}}(\overline D)$ based on the
normal approximation. These intervals treat $\widehat\Sigma_{\rm ref}$ as the
true covariance matrix, so they ignore the simulation error in
$\widehat\Sigma_{\rm ref}$ itself. They may also be inaccurate for the
heavy-tailed GPD, because a few very large squared errors can dominate
$\overline D$.

\subsection{Positive definiteness and Frobenius accuracy}
Table~\ref{tab:nonpd} reports numerical non-PD percentages, and
Table~\ref{tab:rmse} summarizes the RRMSE. Every displayed
method was evaluated on the same $1000$ samples per scenario.
\begin{table}[htbp]\centering\small
\caption{Percentage of the $B=1000$ replications in which each covariance
estimate was classified as numerically non-PD (Section~\ref{sec:pd});
$p=4$. ES: exact unbiased estimator; $\mathrm{ES}_K$: ES after eigenvalue
flooring with condition-number cap $K$; HP: H\'ajek-projection estimator;
JK: delete-one jackknife estimator. A value of 0.0 means that every
estimate was numerically positive definite.}
\label{tab:nonpd}
\begin{tabular}{lrrrrrrr}\toprule
Distribution & $n$ & ES & $\mathrm{ES}_{100}$ & $\mathrm{ES}_{1000}$ & $\mathrm{ES}_{10^5}$ & HP & JK\\\midrule
Normal & 20 & 50.3 & 0.0 & 0.0 & 0.0 & 0.0& 0.0 \\
Normal & 50 & 0.0 & 0.0 & 0.0 & 0.0 & 0.0& 0.0 \\
Gumbel & 20 & 49.9 & 0.0 & 0.0 & 0.0 & 0.0& 0.0 \\
Gumbel & 50 & 0.2 & 0.0 & 0.0 & 0.0 & 0.0& 0.0 \\
GPD$(\xi=0.2)$ & 20 & 59.8 & 0.0 & 0.0 & 0.0 & 0.0& 0.0 \\
GPD$(\xi=0.2)$ & 50 & 7.1 & 0.0 & 0.0 & 0.0 & 0.0& 0.0 \\
\bottomrule\end{tabular}
\end{table}
\begin{table}[htbp]\centering\small
\caption{Relative root mean squared Frobenius error (RRMSE) in
\eqref{eq:rrmse} of each covariance estimator with respect to the
reference covariance matrix; $p=4$ and $B=1000$. Smaller values mean
more accurate estimation. Columns are as in Table~\ref{tab:nonpd}.}
\label{tab:rmse}
\begin{tabular}{lrrrrrrr}\toprule
Distribution & $n$ & ES & $\mathrm{ES}_{100}$ & $\mathrm{ES}_{1000}$ & $\mathrm{ES}_{10^5}$ & HP & JK\\\midrule
Normal & 20 & 0.5064 & 0.5045 & 0.5053 & 0.5054 & 0.4729& 0.5064 \\
Normal & 50 & 0.3070 & 0.3070 & 0.3070 & 0.3070 & 0.3003& 0.3072 \\
Gumbel & 20 & 0.7754 & 0.7737 & 0.7742 & 0.7743 & 0.7105& 0.7759 \\
Gumbel & 50 & 0.4825 & 0.4825 & 0.4825 & 0.4825 & 0.4671& 0.4831 \\
GPD$(\xi=0.2)$ & 20 & 2.5064 & 2.5058 & 2.5057 & 2.5057 & 2.1789& 2.5096 \\
GPD$(\xi=0.2)$ & 50 & 1.4931 & 1.4933 & 1.4931 & 1.4931 & 1.4182& 1.4942 \\
\bottomrule\end{tabular}
\end{table}

At $n=20$, numerical non-PD rates for the ES estimator
were 50.3\%, 49.9\%, and 59.8\% for the normal, Gumbel, and GPD cases,
respectively, whereas no HP or JK estimates were classified as non-PD.
All flooring variants were PD in these simulations by construction.
The HP estimator yielded a smaller RRMSE than the
uncorrected and corrected ES estimators in all six evaluated settings.
For example, in the GPD case with $n=20$, RRMSE decreased from 2.5064
(ES) to 2.1789 (HP). These are empirical findings for the specified
settings, not uniform dominance claims. The JK estimator was also
numerically PD in every replication. Its RRMSE, however, was close to
that of the uncorrected ES estimator and was larger than that of HP
in all six settings.

\subsection{Condition numbers and paired comparison of errors}
Table~\ref{tab:cond} presents median spectral condition numbers. For the
uncorrected ES estimator the median is taken \emph{only over numerically PD
replications}; non-PD matrices have undefined or infinite
condition numbers and are excluded from that conditional median.
Consequently a seemingly moderate value for the uncorrected ES estimator
must be read alongside its non-PD rate. No equivalent exclusion is
required for HP and JK in the reported settings.
\begin{table}[htbp]\centering\small
\caption{Median spectral condition number $\kappa_2$ in \eqref{eq:cond}
over the numerically PD replications; $p=4$ and $B=1000$. Columns ES to JK
are as in Table~\ref{tab:nonpd}. The column Ref gives $\kappa_2(\widehat\Sigma_{\rm ref})$ of the reference
covariance matrix. Values marked $\dagger$ equal the cap $K$; they are
imposed by the flooring correction and do not indicate a better
estimate.}
\label{tab:cond}
\begin{tabular}{lrrrrrrrr}\toprule
Distribution & $n$ & Ref & ES & $\mathrm{ES}_{100}$ & $\mathrm{ES}_{1000}$ & $\mathrm{ES}_{10^5}$ & HP & JK\\\midrule
Normal & 20 & 23.8 & 102.5 & 100.0$^\dagger$ & 1000.0$^\dagger$ & 100000.0$^\dagger$ & 72.5 & 53.8 \\
Normal & 50 & 30.8 & 60.3 & 60.3 & 60.3 & 60.3 & 50.6 & 44.9 \\
Gumbel & 20 & 45.9 & 155.0 & 100.0$^\dagger$ & 1000.0$^\dagger$ & 27555.5 & 97.1 & 74.8 \\
Gumbel & 50 & 57.5 & 95.5 & 95.5 & 95.5 & 95.5 & 77.2 & 69.9 \\
GPD$(\xi=0.2)$ & 20 & 899.7 & 2246.8 & 100.0$^\dagger$ & 1000.0$^\dagger$ & 100000.0$^\dagger$ & 878.3 & 741.6 \\
GPD$(\xi=0.2)$ & 50 & 1204.2 & 1461.0 & 100.0$^\dagger$ & 1000.0$^\dagger$ & 1614.7 & 1073.3 & 1018.1 \\
\bottomrule\end{tabular}
\end{table}
For $n=20$ the uncorrected ES estimator was often not PD; after flooring,
its median condition numbers were at or below the cap. For the
$\mathrm{ES}_K$ columns, Table~\ref{tab:cond} thus only confirms that the
correction works as designed. The cap, however, matters in practice. At
$K=100$, the corrected GPD estimate had a median condition number of 100,
whereas the reference matrix had 899.7 at $n=20$ and 1204.2 at $n=50$.
Such a cap raises the small eigenvalues well above their target values.
This hardly changes the RRMSE in Table~\ref{tab:rmse}, which is dominated
by the large entries, but it can change the inverse matrix substantially.
Hence $K$ should not be smaller than plausible values of
$\kappa_2(\Sigma_n)$, and condition numbers should be compared with the
column Ref rather than ranked by size. The HP and JK estimators require no
such choice.

The JK estimator had smaller median condition numbers than HP in all six
settings; by \eqref{eq:hpjk}, HP shrinks the higher-order components more,
which can spread its eigenvalues. Relative to Ref, both were larger in the
normal and Gumbel cases, with JK closer, and smaller in the GPD case, with
HP closer. This did not give JK a smaller estimation error.

Table~\ref{tab:paired} compares the estimation errors in pairs. For each
replication, $D_b(M)$ in \eqref{eq:paired} is the difference between the
squared Frobenius errors of method $M$ and HP on the same sample.

\begin{table}[htbp]\centering\small
\caption{Mean paired difference $\overline D$ of squared Frobenius errors
in \eqref{eq:paired}: (a) uncorrected ES minus HP; (b) JK minus HP.
Positive values favor HP. Intervals are $\overline D\pm1.96\,\widehat{\mathrm{MCSE}}$ and condition on the
simulated reference covariance.}
\label{tab:paired}
\begin{tabular}{lrrrr}\toprule
Distribution & $n$ & $\overline D$ & MCSE & Approx. 95\% interval\\\midrule
\multicolumn{5}{l}{(a) ES minus HP}\\
Normal & 20 & 8.48e-05 & 3.87e-06 & [7.72e-05, 9.23e-05] \\
Normal & 50 & 1.67e-06 & 6.96e-08 & [1.54e-06, 1.81e-06] \\
Gumbel & 20 & 8.47e-04 & 1.21e-04 & [6.09e-04, 1.08e-03] \\
Gumbel & 50 & 2.06e-05 & 2.46e-06 & [1.58e-05, 2.54e-05] \\
GPD$(\xi=0.2)$ & 20 & 8.51e-02 & 3.50e-02 & [1.65e-02, 1.54e-01] \\
GPD$(\xi=0.2)$ & 50 & 1.90e-03 & 7.04e-04 & [5.19e-04, 3.28e-03] \\\midrule
\multicolumn{5}{l}{(b) JK minus HP}\\
Normal & 20 & 8.49e-05 & 4.20e-06 & [7.67e-05, 9.31e-05] \\
Normal & 50 & 1.72e-06 & 7.90e-08 & [1.57e-06, 1.88e-06] \\
Gumbel & 20 & 8.53e-04 & 1.25e-04 & [6.07e-04, 1.10e-03] \\
Gumbel & 50 & 2.14e-05 & 2.64e-06 & [1.62e-05, 2.66e-05] \\
GPD$(\xi=0.2)$ & 20 & 8.60e-02 & 3.52e-02 & [1.71e-02, 1.55e-01] \\
GPD$(\xi=0.2)$ & 50 & 1.93e-03 & 7.10e-04 & [5.34e-04, 3.32e-03] \\
\bottomrule\end{tabular}
\end{table}
Both panels give a positive $\overline D$ in every setting, so HP had a
smaller mean squared Frobenius error than both the uncorrected ES and the
JK estimators. In the heavy-tailed GPD case, a few large squared errors
can strongly affect $\overline D$, so these approximate intervals should
be interpreted with some caution.

\subsection{Finite-sample bias of the HP and JK variance estimators}
\label{sec:bias}
The theoretical H\'ajek projection has variance no greater
than that of the full U-statistic, by orthogonality of the Hoeffding
components. This does not imply that the empirical HP variance
estimator is always downward biased. For the JK estimator,
Proposition~\ref{prop:bias} gives $\E(\widehat V_{\mathrm{JK},m})\ge\Var(l_m)$,
but it does not give the size of the bias. To examine both estimators,
we used the same $B=1000$ paired samples and the same independent
reference covariance as above. For the $m$th diagonal entry, define
\begin{equation}\label{eq:relbias}
 \operatorname{RB}_{M,m}=
 \frac{\E(\widehat V_{M,m})-V_m}{V_m},\qquad
 \Delta_{M,m}=\frac{\E(\widehat V_{M,m}-\widehat V_{\mathrm{ES},m})}{V_m},
 \quad V_m=\Var(l_m),
\end{equation}
for $M=\mathrm{HP}$ or JK, where $m$ is the order of the L-moment and the
degree of its U-statistic kernel. We estimated $V_m$ by the matching
diagonal entry of $\widehat\Sigma_{\rm ref}$. Since
$\E(\widehat V_{\mathrm{ES},m})=V_m$, we have $\Delta_{M,m}=\operatorname{RB}_{M,m}$.
Table~\ref{tab:biascontrast} reports $100\widehat\Delta_{\mathrm{HP},m}$ and
$100\widehat\Delta_{\mathrm{JK},m}$ with their paired Monte Carlo standard
errors, in percentage points. The paired contrasts have much smaller
Monte Carlo errors than direct estimates of the relative bias. The first
L-moment is omitted because the three estimators agree numerically for
$m=1$.
\begin{table}[htbp]\centering\small
\caption{Paired relative variance contrasts $100\widehat\Delta_{\mathrm{HP},m}$ and
$100\widehat\Delta_{\mathrm{JK},m}$ in percentage points: (a) HP minus ES and
(b) JK minus ES; $B=1000$. Parentheses give paired Monte Carlo standard
errors. Negative values mean that HP or JK gives smaller variance
estimates than the unbiased ES estimator on average, and positive values
mean larger ones. Reference variance uncertainty is not included.}
\label{tab:biascontrast}
\begin{tabular}{lrrrr}\toprule
Distribution & $n$ & $l_2$ & $l_3$ & $l_4$\\\midrule
\multicolumn{5}{l}{(a) HP minus ES}\\
Normal & 20 & $-4.30\;(0.18)$ & $-2.67\;(0.51)$ & $+4.86\;(1.39)$\\
Normal & 50 & $-1.61\;(0.04)$ & $-0.69\;(0.12)$ & $+2.74\;(0.28)$\\
Gumbel & 20 & $-6.07\;(0.27)$ & $-7.08\;(0.83)$ & $-4.90\;(1.92)$\\
Gumbel & 50 & $-2.45\;(0.07)$ & $-2.94\;(0.21)$ & $-1.75\;(0.43)$\\
GPD$(\xi=0.2)$ & 20 & $-10.01\;(0.74)$ & $-17.39\;(2.13)$ & $-21.81\;(4.02)$\\
GPD$(\xi=0.2)$ & 50 & $-3.56\;(0.18)$ & $-5.76\;(0.51)$ & $-6.81\;(0.97)$\\\midrule
\multicolumn{5}{l}{(b) JK minus ES}\\
Normal & 20 & $+6.65\;(0.08)$ & $+21.27\;(0.33)$ & $+46.55\;(0.98)$\\
Normal & 50 & $+2.53\;(0.02)$ & $+7.93\;(0.07)$ & $+16.80\;(0.18)$\\
Gumbel & 20 & $+4.34\;(0.07)$ & $+15.31\;(0.35)$ & $+33.61\;(1.03)$\\
Gumbel & 50 & $+1.58\;(0.01)$ & $+5.49\;(0.07)$ & $+11.74\;(0.20)$\\
GPD$(\xi=0.2)$ & 20 & $+1.29\;(0.03)$ & $+5.99\;(0.25)$ & $+14.63\;(0.94)$\\
GPD$(\xi=0.2)$ & 50 & $+0.48\;(0.01)$ & $+2.15\;(0.05)$ & $+4.91\;(0.15)$\\
\bottomrule\end{tabular}
\end{table}
The HP estimator was smaller on average for most variance components.
The normal $l_4$ result went in the opposite direction at both sample
sizes, where HP exceeded the unbiased ES estimator on average. This is
evidence against a universal downward bias of the empirical HP
estimator, as distinct from the variance of the theoretical H\'ajek
projection. The paired standard errors condition on the simulated
reference variance, and heavy tails make their normal approximations
less reliable.

The JK contrast was positive in every setting, and all the approximate
95\% intervals excluded zero, in agreement with
Proposition~\ref{prop:bias}. The overestimation increased with $m$ and
decreased with $n$. It was largest for $l_4$ at $n=20$, about 47, 34,
and 15 percentage points for the normal, Gumbel, and GPD cases. In the
normal and Gumbel cases, the HP biases of $l_3$ and $l_4$ were much
smaller in magnitude than the JK biases. In the GPD case at both sample
sizes, and for $l_2$ in the Gumbel case, the JK biases were smaller in
magnitude. Hence neither leave-one-out estimator has a smaller bias in
every setting, although HP had the smaller Frobenius error in all
settings (Tables~\ref{tab:rmse} and~\ref{tab:paired}). The remark
following the proof of Proposition~\ref{prop:bias} in
Appendix~\ref{app:asymptotics} discusses the direction of the HP bias
further.

\section{Fast rank-prefix computation of the covariance estimators}\label{sec:fastHP}
The ES, HP, and JK estimators can be computed more efficiently.
The idea is to sort the data once and reuse weighted cumulative sums.
This is a standard computational principle, illustrated by summed-area
tables \citep{Crow1984}. Fast algorithms for rank-based statistics also
avoid repeated pairwise calculations \citep{Christensen2005}.
Here we apply weighted prefix sums to these sample L-moment covariance
estimators, without changing any of them.

For HP and JK, the direct algorithm recomputes each sample L-moment after
removing one observation. Write $x_{1:n}\le\cdots\le x_{n:n}$ and
$l_m=\sum_{j=1}^n w_{m,j}^{(n)}x_{j:n}$. Then
\begin{equation}\label{eq:fastloo}
 l_m^{(-i)}=\sum_{j<i}w_{m,j}^{(n-1)}x_{j:n}
 +\sum_{j>i}w_{m,j-1}^{(n-1)}x_{j:n}.
\end{equation}
The two sums are evaluated from prefix and suffix arrays.
The influence estimates remain $(n-m)(l_m-l_m^{(-i)})$, and the JK
estimator \eqref{eq:jk} uses the same values $l_m^{(-i)}$.

For the ES estimator, the defining double sums have factors
that separate by ordered rank. The identity
\begin{equation}\label{eq:prefixidentity}
\sum_{1\le i<j\le n}a_i b_j
=\sum_{j=2}^n b_j\left(\sum_{i=1}^{j-1}a_i\right)
\end{equation}
replaces a quadratic pairwise sum with one running cumulative sum.
For fixed $p$, the direct ES implementation has $O(p^2n^2)$
post-sorting work, whereas the prefix implementation has $O(p^2n)$.
For HP and JK, sorting, leave-one-out contributions, and forming the covariance
matrix require $O(n\log n+np^2)$ work.
Appendix~\ref{app:fast} gives elementary derivations and pseudocode.

The direct and fast implementations of each estimator agreed to
numerical precision. Table~\ref{tab:fast-bench} compares elapsed times per call.
All six columns were obtained in one run on the same samples.
Each fast method was called $10{,}000$ times per sample size. Each direct
method was called 100, 100, 60, 12, and 4 times for $n=25$, 50, 100, 500,
and 1000, respectively.

\begin{table}[htbp]\centering\small
\caption{Elapsed computation time (ms per call) for covariance estimation
($p=4$). Fast methods were timed together using 10,000 repetitions
per sample size; direct methods were measured separately.}
\label{tab:fast-bench}
\begin{tabular}{rrrrrrrr}\toprule
$n$ & Direct ES & Fast ES & Direct HP & Fast HP & Direct JK & Fast JK & Fast ES/fast HP \\\midrule
25 & 7.50 & 1.161 & 6.80 & 0.976 & 8.10 & 0.827 & 1.19 \\
50 & 15.30 & 1.280 & 13.80 & 0.878 & 13.80 & 0.832 & 1.46 \\
100 & 38.00 & 1.437 & 31.50 & 0.864 & 29.67 & 0.914 & 1.66 \\
500 & 162.50 & 2.967 & 210.83 & 1.294 & 217.50 & 1.318 & 2.29 \\
1000 & 390.00 & 4.823 & 602.50 & 1.861 & 597.50 & 1.814 & 2.59 \\
\bottomrule\end{tabular}
\end{table}
The fast HP implementation was about 1.2--2.6 times faster than the
fast ES implementation in the matched benchmark. The fast JK and fast
HP implementations had essentially the same computing time, as expected
from their common leave-one-out computation. The rank-prefix
algorithms substantially reduced the cost of the respective direct
implementations.

\paragraph{Comparison with the bootstrap.}
For comparison, we also timed the nonparametric bootstrap estimate of the
covariance matrix on the data used for Table~\ref{tab:fast-bench}.
For $n=25$ to 1000, the bootstrap took 74--350 ms per call with $B=1000$
resamples and 9--28 ms per call with $B=100$ resamples. These times are
about 70--170 times and 9--14 times, respectively, the time of the fast
HP implementation. For confidence intervals or regions in small samples,
however, a bootstrap approach may be preferable, because the bootstrap
distribution can reflect the skewness of the sampling distribution of the
L-moment ratios, which a normal approximation based on a covariance
estimate ignores \citep{Hosking2015}.

\section{Discussion}\label{sec:discussion}
\subsection{Limitations}
None of the three covariance estimators considered here is new. The ES
estimator was derived by \citet{ElamirSeheult2004}, the HP estimator
coincides with the structural-components estimator of \citet{Sen1960},
and the JK estimator is the classical delete-one jackknife. The
contribution of this paper is to apply the HP and JK estimators to sample
L-moments, to relate them exactly through \eqref{eq:hpjk}, and to compare
them with the ES estimator and its eigenvalue-floored versions.

The study also has several limitations. First, the HP and JK estimators
are positive semidefinite, but they are not necessarily positive
definite. They can be singular, for example, when a small sample contains
many tied values, so the rank and the condition number should still be
checked before inversion. Second, neither estimator is unbiased. The JK
variance estimates are conservative (Proposition~\ref{prop:bias}),
whereas the HP variance estimates were negatively biased in the
heavy-tailed GPD case. Third, the simulation covered only $p=4$, three
distributions, and two sample sizes, and the reference covariance
matrices were themselves estimated by Monte Carlo simulation. Their
split-half relative differences ranged from 0.89\% to 2.77\%. This
reference uncertainty was not propagated into the reported RRMSEs, so
small differences at $n=50$ should be interpreted with caution. Finally,
the theory requires $\E(X^2)<\infty$, which excludes, for example, the GPD
with $\xi\ge1/2$.

\subsection{Extensions to other L-moment-type statistics}
The construction in Section~\ref{sec:estimators} uses only the fact that
$l_m$ is a U-statistic of degree $m$. It therefore extends to other
U-statistics. Let $U$ be a U-statistic of degree $k$ with a
square-integrable kernel, and let $U^{(-i)}$ be its value after $X_i$ is
removed. The subset-count identity \eqref{eq:subsetcount} becomes
$U=(k/n)A_i+\{(n-k)/n\}U^{(-i)}$, where $A_i$ is the average of the kernel
over the subsets of size $k$ that contain $i$. Hence the empirical
influence value is $\widehat\phi_i=(n-k)(U-U^{(-i)})$. For a vector of such
statistics with kernel degrees $k_1,\ldots,k_p$, the relation
\eqref{eq:hpjk} holds with
\[
D_n=\operatorname{diag}\Bigl(\frac{n-k_1}{n-1},\ldots,\frac{n-k_p}{n-1}\Bigr).
\]
Examples are the trimmed L-moments of \citet{ElamirSeheult2003}, whose
sample version of order $m$ with trimming $(t_1,t_2)$ is a U-statistic of
degree $m+t_1+t_2$, and the LH-moments of \citet{Wang1997} \citep[see also][]{SeenoiEtAl2026LHmom}, whose sample
version of order $m$ and level $\eta$ is a U-statistic of degree $m+\eta$.

\subsection{Covariance of parameter and return-level estimators}
Many L-moment estimators of distribution parameters are smooth functions
of the sample L-moments. Let $\boldsymbol\theta=G(\boldsymbol\lambda)\in\mathbb R^r$,
where $G$ is continuously differentiable near $\boldsymbol\lambda$ with
Jacobian matrix $J=\partial G(\boldsymbol\lambda)/\partial\boldsymbol\lambda^\top$,
and let $\widehat{\boldsymbol\theta}=G(\boldsymbol l)$. By the delta method and
\eqref{eq:asylin},
\begin{equation}\label{eq:deltaasy}
\sqrt n(\widehat{\boldsymbol\theta}-\boldsymbol\theta)\xrightarrow{d}
N_r(\boldsymbol0,J\Omega J^\top),
\end{equation}
and a finite-sample approximation is
$\Cov(\widehat{\boldsymbol\theta})\approx J\Sigma_nJ^\top$. Replacing $J$ by
$\widehat J=\partial G(\boldsymbol l)/\partial\boldsymbol\lambda^\top$ and $\Sigma_n$ by
one of the three estimators gives
\begin{equation}\label{eq:deltaest}
\widehat{\Cov}_M(\widehat{\boldsymbol\theta})=\widehat J\,\widehat\Sigma_M\,\widehat J^\top,
\qquad M=\mathrm{ES},\ \mathrm{HP},\ \mathrm{JK}.
\end{equation}
For $M=\mathrm{HP}$ or JK, \eqref{eq:deltaest} is positive semidefinite for
every sample, because
$a^\top\widehat J\widehat\Sigma_M\widehat J^\top a
=(\widehat J^\top a)^\top\widehat\Sigma_M(\widehat J^\top a)\ge0$ for every
$a\in\mathbb R^r$. For $M=\mathrm{ES}$, an indefinite $\widehat\Sigma_{\mathrm{ES}}$
can give a negative variance estimate for a component of
$\widehat{\boldsymbol\theta}$.

The same argument applies to a return level. Let $Q(u;\boldsymbol\theta)$ be the
quantile function, and let $x_T=Q(1-1/T;\boldsymbol\theta)$ be the $T$-year
return level, estimated by $\widehat x_T=Q(1-1/T;\widehat{\boldsymbol\theta})$. Then
\begin{equation}\label{eq:deltareturn}
\Var(\widehat x_T)\approx\boldsymbol c^\top\Sigma_n\boldsymbol c,\qquad
\boldsymbol c=J^\top\nabla_{\boldsymbol\theta}Q(1-1/T;\boldsymbol\theta),
\end{equation}
and the estimate $\widehat{\boldsymbol c}^\top\widehat\Sigma_M\widehat{\boldsymbol c}$ is
nonnegative for $M=\mathrm{HP}$ or JK.

A simple example is the \emph{L-moment ratio}
$t_r=l_r/l_2$, $r=3,4$. Here $G(\boldsymbol\lambda)=\lambda_r/\lambda_2$, whose
gradient has entries $-\lambda_r/\lambda_2^2$ and $1/\lambda_2$ in positions
2 and $r$, and zeros elsewhere. Hence (see also \citet[Eq.~(7)]{Hosking2015})
\begin{equation}\label{eq:deltaratio}
\Var(t_r)\approx\frac{1}{\lambda_2^2}
\bigl\{V_r-2\tau_rC_{2r}+\tau_r^2V_2\bigr\},
\qquad C_{2r}=\Cov(l_2,l_r).
\end{equation}
Since the GEV shape estimator of \citet{HoskingWallisWood1985} is a
smooth function of $t_3$, its approximate variance follows from
\eqref{eq:deltaratio} by one more application of the delta method. These
approximations are of first order. Whether the use of a finite-sample
covariance estimate in \eqref{eq:deltaest}, instead of a plug-in
asymptotic formula, improves accuracy in small samples remains to be
studied.

\subsection{Bias reduction}
By \eqref{eq:hpjk}, the HP and JK estimators are the end points of the
family
\[
\widehat\Sigma(\alpha)=D_n^{\alpha}\,\widehat\Sigma_{\mathrm{JK}}\,D_n^{\alpha},
\qquad 0\le\alpha\le1,
\]
with $\widehat\Sigma(0)=\widehat\Sigma_{\mathrm{JK}}$ and
$\widehat\Sigma(1)=\widehat\Sigma_{\mathrm{HP}}$. Every member of this family is
positive semidefinite. The JK estimator overestimated the higher-order
variances in our simulations, whereas HP underestimated them in the
heavy-tailed case. An intermediate value of $\alpha$, possibly chosen from
the data, might therefore reduce the bias while keeping positive
semidefiniteness. This is left for future work.

\section{Conclusion}\label{sec:conclusion}
We studied numerical definiteness and finite-sample accuracy of
covariance estimators for sample L-moments. The HP estimator avoids
negative eigenvalues by construction and performed well in the simulated
settings. It is an exact diagonal rescaling of the delete-one jackknife
estimator, and it had smaller Frobenius estimation error than the
jackknife estimator in all simulated settings. Eigenvalue flooring
provides an alternative when positive definiteness is required, provided
that the cap $K$ is not smaller than plausible values of the condition
number of $\Sigma_n$. The ES, HP, and JK estimators have the same
first-order asymptotic covariance limit, but their finite-sample
performance differs. Rank-prefix algorithms substantially reduce the
computation time for the three estimators.

\appendix
\section*{Appendix}
\section{Numerical definiteness and eigenvalue correction}\label{app:alg}
This appendix gives the two procedures of Section~\ref{sec:pd} in
algorithmic form. Algorithm~\ref{alg:check} classifies a covariance
estimate as numerically non-PSD, near-singular, or numerically PD, using
the tolerance $\tau$ in \eqref{eq:tol}, and returns its condition number
when the estimate is numerically PD. Algorithm~\ref{alg:floor} computes the
eigenvalue-flooring correction \eqref{eq:floor} with the threshold
\eqref{eq:delta}, so that the corrected matrix is PD and has condition
number at most $K$. It gives the estimates $\mathrm{ES}_K$ in
Section~\ref{sec:sim}.

\begin{algorithm}[htbp]
\caption{Numerical definiteness assessment for a symmetric covariance estimate}
\label{alg:check}
\begin{algorithmic}[1]
\Require Real $p\times p$ matrix $A$ with finite entries;
relative tolerance $\varepsilon_{\rm rel}=10^{-10}$
\State $S\gets(A+A^\top)/2$
\State Compute eigenvalues $d_1,\ldots,d_p$ of $S$
\State $d_{\min}\gets\min_jd_j$; $d_{\max}\gets\max_jd_j$
\State $\tau\gets\varepsilon_{\rm rel}\max(\max_j|d_j|,\epsilon_{\rm mach})$
\If{$d_{\min}<-\tau$}
 \State $\mathrm{status}\gets\texttt{nonPSD}$; $\kappa\gets\infty$
\ElsIf{$d_{\min}\le\tau$}
 \State $\mathrm{status}\gets\texttt{nearSingular}$; $\kappa\gets\infty$
\Else
 \State $\mathrm{status}\gets\texttt{numericallyPD}$;
 $\kappa\gets d_{\max}/d_{\min}$
\EndIf
\State \Return $\mathrm{status},d_{\min},d_{\max},\tau,\kappa$
\end{algorithmic}
\end{algorithm}
\begin{algorithm}[htbp]
\caption{Eigenvalue-flooring correction with a spectral condition-number cap}
\label{alg:floor}
\begin{algorithmic}[1]
\Require Real symmetric estimate $A$ with finite entries; $K>1$ (default $K=10^3$)
\State $S\gets(A+A^\top)/2$; compute $S=Q\operatorname{diag}(d)Q^\top$
\State $s\gets\max_j|d_j|$; if $s=0$, set $s\gets1$
\State $u\gets\max(\max_j d_j,s/K)$; $\delta\gets u/K$
\State $d_j^*\gets\max(d_j,\delta)$ for $j=1,\ldots,p$
\State $S^*\gets Q\operatorname{diag}(d^*)Q^\top$
\State \Return $(S^*+S^{*\top})/2$
\end{algorithmic}
\end{algorithm}
\begin{remark}
Both algorithms first replace $A$ by $(A+A^\top)/2$. A covariance estimate
is symmetric in exact arithmetic, but rounding errors can make the
computed matrix slightly asymmetric; this symmetrization removes such asymmetry
before the eigenvalues are computed. In applications, positive
definiteness is often checked by attempting a Cholesky factorization,
which succeeds only for a positive definite matrix. A successful
factorization, however, does not show that the matrix is well conditioned,
so the condition number should still be examined before inversion.
\end{remark}

\section{Supplementary derivations for Section~\ref{sec:estimators}}\label{app:supp}
This appendix gives the derivations omitted from
Sections~\ref{sec:hp} and~\ref{sec:hpest}.

\subsection{The contribution and the H\'ajek projection}\label{app:suppproj}
The second equality in \eqref{eq:contribdef} holds because a fraction $m/n$
of the subsets in \eqref{eq:ustat} contain $i$. Each kernel value from such
a subset has conditional expectation $\lambda_m+g_m(X_i)$ given $X_i$, and
each kernel value from a subset not containing $i$ has conditional
expectation $\lambda_m$. Hence
$\E(l_m-\lambda_m\mid X_i)=(m/n)g_m(X_i)$.

To show that $H_{n,m}$ in \eqref{eq:hajekprojection} is the $L^2$
projection, note that $\E(c_{m,j}\mid X_i)=\E\{c_{m,j}\}=0$ for $j\ne i$, so
that $\E(H_{n,m}\mid X_i)=c_{m,i}=\E(l_m-\lambda_m\mid X_i)$ by
\eqref{eq:contribdef}. Therefore
\[
\E\{(l_m-\lambda_m-H_{n,m})\,a(X_i)\}
=\E\bigl[\E(l_m-\lambda_m-H_{n,m}\mid X_i)\,a(X_i)\bigr]=0,
\qquad i=1,\ldots,n,
\]
for every square-integrable function $a$. Thus $l_m-\lambda_m-H_{n,m}$ is
orthogonal to every sum $\sum_{i=1}^na_i(X_i)$, and $H_{n,m}$ is itself
such a sum.

\subsection{The influence function}\label{app:suppif}
The second equality in \eqref{eq:ifderiv} holds because $\lambda_m(F)$ is
an $m$-fold integral of $h_m$ with respect to $F$. Differentiating the
$m$-fold product measure of $(1-\varepsilon)F+\varepsilon\Delta_x$ at
$\varepsilon=0$ replaces one of its $m$ factors $F$ by $\Delta_x-F$ in each
of the $m$ first-order terms. Each such term equals
$\E\{h_m(x,X_2,\ldots,X_m)\}-\lambda_m=g_m(x)$ by the symmetry of $h_m$, so
the derivative is $m\,g_m(x)=\phi_m(x)$.

\subsection{Empirical influence values}\label{app:suppemp}
For \eqref{eq:condexp}, each subset that contains $i$ gives a kernel value
with conditional expectation $\lambda_m+g_m(X_i)$ given $X_i$, so
$\E(A_{m,i}\mid X_i)=\lambda_m+g_m(X_i)$. By \eqref{eq:contribdef},
$\E(l_m\mid X_i)=\lambda_m+(m/n)g_m(X_i)$. Hence, by
\eqref{eq:contribution},
\[
\E(\widehat\phi_{m,i}\mid X_i)
=m\Bigl(1-\frac mn\Bigr)g_m(X_i)=\frac{n-m}{n}\,\phi_m(X_i).
\]

For \eqref{eq:center}, each of the $\binom nm$ subsets in \eqref{eq:ustat}
contains exactly $m$ observations, so it is counted $m$ times in
$\sum_{i=1}^n A_{m,i}$. Since each $A_{m,i}$ is an average over
$\binom{n-1}{m-1}$ subsets,
\[
\sum_{i=1}^n A_{m,i}
=\frac{m\binom nm}{\binom{n-1}{m-1}}\,l_m=n\,l_m .
\]
Thus, by \eqref{eq:contribution},
$\sum_{i=1}^n\widehat\phi_{m,i}=m\bigl(\sum_{i=1}^nA_{m,i}-n\,l_m\bigr)=0$.

\section{Proof of propositions}\label{app:asymptotics}
This appendix proves Propositions~\ref{prop:bias} and~\ref{prop:equiv}.
Throughout, the observations are i.i.d.\ with $\E(X^2)<\infty$, and $p$ is
fixed. Since each kernel $h_m$ is a linear combination of order
statistics of a fixed number of observations, these conditions ensure
that the kernels and their pairwise products have finite expectations.

\subsection{Proof of Proposition~\ref{prop:bias}}\label{app:hpjackknife}
\begin{proof}
Fix $m\le p$. Let $S$ denote the sample L-moment of order $m$ computed
from $n-1$ observations, and let $V_{n-1,m}=\Var(S)$. For the leave-one-out
values, $S_{(i)}=l_m^{(-i)}$, and by \eqref{eq:jk},
$\widehat V_{\mathrm{JK},m}=\{(n-1)/n\}\sum_{i=1}^n(S_{(i)}-\bar S_{(\cdot)})^2$.
Since $S$ is a symmetric statistic of $n-1$ i.i.d.\ observations with
finite variance, the inequality of \citet{EfronStein1981} gives
\[
\E(\widehat V_{\mathrm{JK},m})
=\frac{n-1}{n}\,\E\Bigl\{\sum_{i=1}^n(S_{(i)}-\bar S_{(\cdot)})^2\Bigr\}
\ge\frac{n-1}{n}\,V_{n-1,m}.
\]

Next, $S$ and $l_m$ are U-statistics of degree $m$ with the same kernel,
based on $n-1$ and $n$ observations. For such U-statistics, $n$ times the
variance is nonincreasing in the sample size
\citep[][Lemma~5.2.1A]{Hoeffding1948,Serfling1980}, so that
$(n-1)V_{n-1,m}\ge n\Var(l_m)$. Combining the two inequalities gives
$\E(\widehat V_{\mathrm{JK},m})\ge\Var(l_m)$.

The bound for the HP estimator
follows from \eqref{eq:jkdiag}, because
$\E(\widehat V_{\mathrm{HP},m})=\{(n-m)/(n-1)\}^2\E(\widehat V_{\mathrm{JK},m})$.
\end{proof}

\begin{remark}
The inequality \eqref{eq:jkdiag} holds for each observed sample. It compares
the two estimates with each other, but it says nothing directly about
their bias relative to $\Var(l_m)$. For the HP estimator, the bound in
Proposition~\ref{prop:bias} does not determine the sign of the bias. The
orthogonal Hoeffding decomposition implies $\Var(H_{n,m})\le\Var(l_m)$, but
$\widehat V_{\mathrm{HP},m}$ is an estimated first-order variance, not
$\Var(H_{n,m})$ itself. We therefore assess the finite-sample bias of the HP
estimator empirically in Section~\ref{sec:bias}.
\end{remark}

\subsection{Proof of Proposition~\ref{prop:equiv}}
The Hoeffding decomposition gives the limit of the target covariance,
$n\Sigma_n\to\Omega$. Indeed, writing $\boldsymbol R_n$ for the vector of
higher-order terms of the Hoeffding decomposition,
\[
\boldsymbol l-\boldsymbol\lambda
=\frac1n\sum_{i=1}^n\boldsymbol\phi(X_i)+\boldsymbol R_n,
\qquad \E\|\boldsymbol R_n\|^2=O(n^{-2}),
\]
and the orthogonality of the Hoeffding components implies
$\Sigma_n=n^{-1}\Omega+O(n^{-2})$ entrywise. This does not by itself show
that the covariance estimators converge to the same limit. We prove this
in two lemmas.

\begin{lemma}\label{lem:es}
Under the conditions of Proposition~\ref{prop:equiv},
$n\widehat\Sigma_{\mathrm{ES}}\xrightarrow{p}\Omega$.
\end{lemma}
\begin{proof}
Fix orders $m,q\le p$ and set $r=\min(m,q)$. For $0\le s\le r$, define
\[
 \theta_s^{m,q}=\E\bigl\{h_m(X_A)h_q(X_B)\bigr\},
 \qquad |A|=m,\quad |B|=q,\quad |A\cap B|=s,
\]
where the expectation is over $m+q-s$ independent variables. Thus
$\theta_0^{m,q}=\lambda_m\lambda_q$ and
$\theta_1^{m,q}-\theta_0^{m,q}=\E\{g_m(X)g_q(X)\}$.

For independently drawn index sets $A,B$ of sizes $m,q$ from
$\{1,\ldots,n\}$, the overlap probability is
\[
 \pi_{n,s}^{m,q}
 =\frac{\binom ms\binom{n-m}{q-s}}{\binom nq}.
\]
Conditioning on $|A\cap B|$ gives the exact identity
\begin{equation}\label{eq:overlapcov}
 \operatorname{Cov}(l_m,l_q)
 =\sum_{s=0}^{r}\bigl(\pi_{n,s}^{m,q}-\mathbf 1\{s=0\}\bigr)
 \theta_s^{m,q}.
\end{equation}

Let $\widehat\theta_s^{m,q}$ be the complete U-statistic obtained by
symmetrizing the product $h_m(X_A)h_q(X_B)$ over all distinct unions
$A\cup B$ of size $m+q-s$. In particular,
$\widehat\theta_0^{m,q}$ is the disjoint-subset estimator of
$\lambda_m\lambda_q$, which is available when $n\ge m+q$; the condition
$n\ge2p$ ensures this for all $m,q\le p$.
Replacing each $\theta_s^{m,q}$ in \eqref{eq:overlapcov} with its
unbiased U-statistic gives
\begin{equation}\label{eq:unbiasedoverlap}
 \widehat\Sigma_{\mathrm{ES},mq}
 =\sum_{s=0}^{r}\bigl(\pi_{n,s}^{m,q}-\mathbf 1\{s=0\}\bigr)
 \widehat\theta_s^{m,q},
\end{equation}
which is an equivalent representation of the ES estimator; see
Remark~\ref{rem:unique} below.

For fixed $m,q$, the law of large numbers
for integrable fixed-degree U-statistics yields
$\widehat\theta_s^{m,q}\xrightarrow{p}\theta_s^{m,q}$.
As $n\to\infty$, the overlap coefficients obey
\[
 n(\pi_{n,0}^{m,q}-1)\longrightarrow -mq,\qquad
 n\pi_{n,1}^{m,q}\longrightarrow mq,\qquad
 n\pi_{n,s}^{m,q}\longrightarrow0\quad(s\ge2).
\]
Multiplying \eqref{eq:unbiasedoverlap} by $n$ and taking limits gives
\[
 n\widehat\Sigma_{\mathrm{ES},mq}\xrightarrow{p}
 mq(\theta_1^{m,q}-\theta_0^{m,q})
 =\E\{\phi_m(X)\phi_q(X)\}=\Omega_{mq}.
\]
\end{proof}

\begin{lemma}\label{lem:jk}
Under the conditions of Proposition~\ref{prop:equiv},
$n\widehat\Sigma_{\mathrm{JK}}\xrightarrow{p}\Omega$ and
$n\widehat\Sigma_{\mathrm{HP}}\xrightarrow{p}\Omega$.
\end{lemma}
\begin{proof}
For a U-statistic of fixed degree with a square-integrable kernel, $n$ times
its delete-one jackknife variance estimate converges in probability to the
limit of $n$ times its variance; this is Theorem~6 of \citet{Arvesen1969}
with $g$ the identity function, and it requires only a finite second
moment of the kernel.

Fix $a\in\mathbb R^p$. The linear combination
$a^\top\boldsymbol l$ is again a U-statistic of degree at most $p$ with a
square-integrable kernel, because a U-statistic of degree $m\le p$ can be
rewritten as one of degree $p$ by averaging its kernel over subsets. Since
the jackknife is linear in the leave-one-out values,
$a^\top\widehat\Sigma_{\mathrm{JK}}a$ is the jackknife variance estimate of
$a^\top\boldsymbol l$. Hence
$n\,a^\top\widehat\Sigma_{\mathrm{JK}}a\xrightarrow{p}a^\top\Omega a$. As this
holds for every $a$, the polarization identity
$2a^\top Mb=(a+b)^\top M(a+b)-a^\top Ma-b^\top Mb$ gives
$n\widehat\Sigma_{\mathrm{JK}}\xrightarrow{p}\Omega$ entrywise.

The HP limit
then follows from \eqref{eq:hpjk}, because
$n\widehat\Sigma_{\mathrm{HP}}=D_n(n\widehat\Sigma_{\mathrm{JK}})D_n$ and
$D_n\to I_p$ for fixed $p$.
\end{proof}

\begin{proof}[Proof of Proposition~\ref{prop:equiv}]
Lemmas~\ref{lem:es} and~\ref{lem:jk} give the three limits in
\eqref{eq:equiv}. Subtracting the probability limits gives the three
statements on the differences. Entrywise convergence of $p\times p$
matrices implies convergence in every matrix norm.
\end{proof}

\begin{remark}\label{rem:unique}
The symmetric distribution-free unbiased estimator is unique on the class
of all i.i.d.\ laws (where expectations exist), so
\eqref{eq:unbiasedoverlap} and the ES estimator coincide. To see
uniqueness, if two symmetric statistics of fixed sample size have zero
difference in expectation for every finitely supported distribution,
expand that expectation as a polynomial in its point masses; its
coefficients for each sample multiplicity must vanish. Consequently the
statistics agree on every sample, up to the usual null-set conventions.
This gives a representation of the ES estimator without reproducing its
order-statistic coefficient formula.
\end{remark}

\section{Rank-prefix algorithms for covariance estimation}\label{app:fast}
The algorithms below compute exactly the same estimators as the direct
formulas, but in a faster way. After the data are sorted once, every sum
that the direct formulas evaluate repeatedly is obtained from cumulative
sums over the ranks, so that the work after sorting is linear in $n$ for
fixed $p$.

\subsection{Fast ES covariance by weighted prefix sums}
Let $b_k$ and $(a)_{(r)}$ be as in \eqref{eq:pwm}, and let $C$ be the
matrix in \eqref{eq:lCb}.
The formula \eqref{eq:esb} of \citet{ElamirSeheult2004} can be written as
\begin{align}
\widehat{\Cov}(b_k,b_l)
 &=b_kb_l-\widehat B_{kl},\label{eq:app-elamir}\\
\widehat B_{kl}
 &=\frac{1}{(n)_{(k+l+2)}}\sum_{1\le i<j\le n}x_{i:n}x_{j:n}\nonumber\\
 &\quad\times\big\{(i-1)_{(k)}(j-k-2)_{(l)}
             +(i-1)_{(l)}(j-l-2)_{(k)}\big\}.\nonumber
\end{align}
Here $0\le k,l\le p-1$ and $n\ge2p$. Define
\begin{equation}\label{eq:app-prefix}
P_k(j)=\sum_{i=1}^{j}(i-1)_{(k)}x_{i:n},\qquad P_k(0)=0.
\end{equation}
The first numerator term of \eqref{eq:app-elamir} becomes
\[
\sum_{j=2}^n x_{j:n}(j-k-2)_{(l)}P_k(j-1).
\]
The second term exchanges $k$ and $l$. Thus
\begin{equation}\label{eq:app-elamir-fast}
\widehat B_{kl}=\frac{1}{(n)_{(k+l+2)}}\sum_{j=2}^n x_{j:n}
\big\{(j-k-2)_{(l)}P_k(j-1)
 +(j-l-2)_{(k)}P_l(j-1)\big\}.
\end{equation}

Here $P_k(j)$ is the cumulative (prefix) sum of the weighted order
statistics $(i-1)_{(k)}x_{i:n}$ up to rank $j$. Computing the prefix sums
$P_k(1),\ldots,P_k(n)$ in \eqref{eq:app-prefix} for $k=0,\ldots,p-1$ costs
$O(np)$, and evaluating \eqref{eq:app-elamir-fast} for all pairs $(k,l)$
costs $O(np^2)$. The PWM-to-L-moment transformation \eqref{eq:es},
$\widehat\Sigma_{\mathrm{ES}}=C\widehat S_bC^\top$, then gives the $p\times p$
covariance matrix of the sample L-moments.

\begin{algorithm}[htbp]
\caption{Fast ES covariance for the first $p$ sample L-moments}
\label{alg:fast-elamir}
\begin{algorithmic}[1]
\Require Observations $x_1,\ldots,x_n$, $n\ge2p$
\State Sort the observations as $x_{1:n},\ldots,x_{n:n}$.
\For{$k=0,\ldots,p-1$}
  \State Compute $P_k(0)=0$ and $P_k(j)=P_k(j-1)+(j-1)_{(k)}x_{j:n}$.
  \State Compute $b_k=(n)_{(k+1)}^{-1}P_k(n)$.
\EndFor
\For{$0\le k\le l\le p-1$}
  \State Compute $\widehat B_{kl}$ using \eqref{eq:app-elamir-fast}.
  \State Set $A_{kl}=A_{lk}=b_kb_l-\widehat B_{kl}$.
\EndFor
\State Form the PWM-to-L-moment transformation matrix $C$ in \eqref{eq:lCb}.
\State \Return $\widehat\Sigma_{\mathrm{ES}}=CAC^\top$.
\end{algorithmic}
\end{algorithm}

\subsection{Fast HP covariance by prefix and suffix sums}
For sample size $N$, let $w_{m,j}^{(N)}$ be the linear
order-statistic weights of the sample L-moment $l_m$, so that
\[
l_m=\sum_{j=1}^{N}w_{m,j}^{(N)}x_{j:N},\qquad
w_{m,j}^{(N)}=\sum_{k=0}^{m-1}C_{m,k+1}\frac{(j-1)_{(k)}}{N\,(N-1)_{(k)}},
\]
which follows from \eqref{eq:pwm} and \eqref{eq:lCb}.
Fix $m$ and use weights at sample size $n-1$. Precompute
\[
L_m(i)=\sum_{j=1}^{i-1}w_{m,j}^{(n-1)}x_{j:n},\qquad
R_m(i)=\sum_{j=i+1}^{n}w_{m,j-1}^{(n-1)}x_{j:n}.
\]
The array $L_m(i)$ can be built from left to right and $R_m(i)$
from right to left. After removing rank $i$, the order of observations
below rank $i$ is unchanged; ranks above $i$ decrease by one.
Consequently $l_m^{(-i)}=L_m(i)+R_m(i)$, which is
\eqref{eq:fastloo}. Using the ordinary sample $l_m$,
\[
\widehat\phi_{m,i}=(n-m)\{l_m-l_m^{(-i)}\}.
\]

Collecting the $p$ values for each rank $i$ into row $i$ of
$\widehat\Phi$, the HP covariance is
\[
\widehat\Sigma_{\mathrm{HP}}=\frac{1}{n(n-1)}\widehat\Phi^\top\widehat\Phi.
\]
This matrix is $\{n(n-1)\}^{-1}$ times the matrix of inner products of the
columns of $\widehat\Phi$, so it is positive semidefinite, as shown in
\eqref{eq:hppsd}. The JK estimator uses the same leave-one-out values $l_m^{(-i)}$. It can be
computed directly from \eqref{eq:jk}, or as
$\widehat\Sigma_{\mathrm{JK}}=D_n^{-1}\widehat\Sigma_{\mathrm{HP}}D_n^{-1}$ by
\eqref{eq:hpjk}; its cost is the same as that of the HP estimator.

\begin{algorithm}[htbp]
\caption{Fast H\'ajek-projection covariance by rank-prefix and suffix sums}
\label{alg:fast-hp}
\begin{algorithmic}[1]
\Require Observations $x_1,\ldots,x_n$ and L-moment orders $1,\ldots,p$, $n>p$
\State Sort the observations once and compute $l_1,\ldots,l_p$.
\For{$m=1,\ldots,p$}
  \State Obtain the $n-1$ observation L-moment weights $w_{m,j}^{(n-1)}$.
  \State Construct $L_m(i)$ by a forward cumulative sum.
  \State Construct $R_m(i)$ by a backward cumulative sum.
  \For{$i=1,\ldots,n$}
    \State Set $\widehat\phi_{m,i}=(n-m)\{l_m-L_m(i)-R_m(i)\}$.
  \EndFor
\EndFor
\State Form the $n\times p$ matrix $\widehat\Phi=(\widehat\phi_{m,i})_{i,m}$.
\State \Return $\widehat\Sigma_{\mathrm{HP}}=\widehat\Phi^\top\widehat\Phi/[n(n-1)]$ and, if
required, $\widehat\Sigma_{\mathrm{JK}}=D_n^{-1}\widehat\Sigma_{\mathrm{HP}}D_n^{-1}$.
\end{algorithmic}
\end{algorithm}

The implementation works with tied values as well: choosing any
consistent ordering among tied observations leaves the resulting
L-moments and leave-one-out values unchanged. With fixed $p$, both
methods need only $O(np)$ cumulative-sum storage if the arrays are
retained. The total HP cost is $O(n\log n+np^2)$, and the same bound
applies to fast ES after sorting. Constant factors depend on the implementation.

\section*{Statements and declarations}
\noindent\textbf{Funding.}
P.S. acknowledges research support from Khon Kaen University, Thailand.
J.P. has received support from the BK21 FOUR program
(No.~5120200913674), funded by the Ministry of Education, Korea,
and the National Research Foundation of Korea.
Jihong Park's work was supported by the National Research Foundation of
Korea (NRF) grants funded by the Korea government (MSIT),
No.~RS-2026-25499038.

\medskip\noindent\textbf{Competing interests.}
The authors declare no competing interests.

\medskip\noindent\textbf{Data and code availability.}
 \textsf{R} code for the numerical experiments and covariance computation is available at \url{https://github.com/Pjihong/sample_Lmom}.

\medskip\noindent\textbf{Use of generative artificial intelligence.}
ChatGPT (OpenAI) and Claude (Anthropic) were used to assist with mathematical exposition,
proof development, R code drafting, and manuscript preparation.
All reported numerical results were obtained from R simulations conducted by the authors.
The authors are responsible for checking the mathematical arguments,
code, citations, results, and final text.

\end{document}